\documentclass[11pt]{article}             
\usepackage{osid}
\usepackage{graphicx}
\usepackage{amsfonts}                    %
\newcommand{\bra}[1]{\langle#1|}

\newcommand{\modul}[1]{|#1|}
\newcommand{\scal}[2]{\langle#1|#2\rangle}\newcommand{\ket}[1]{|#1\rangle}

\title{Generation of Entangled Two-Photons Binomial States\\ in Two Spatially Separate Cavities}
\author{Rosario Lo Franco\thanks{\,\,\,E-mail: lofranco@fisica.unipa.it; URL: http://www.fisica.unipa.it/\~{}lofranco},
Giuseppe Compagno, Antonino Messina, and Anna
Napoli\\{\footnotesize\it INFM, MIUR and Dipartimento di Scienze
Fisiche ed Astronomiche, Universit\'a di Palermo,\break via
Archirafi 36, 90123 Palermo, Italy}}
\date{\today}

\begin{document}

\maketitle
\begin{abstract}
     We propose a conditional scheme to generate entangled two-photons generalized binomial states inside two
     separate single-mode high-$Q$ cavities. This scheme requires that the two cavities are initially prepared in entangled one-photon
     generalized binomial states and exploits the passage of two appropriately prepared two-level atoms one in each
     cavity. The measurement of the ground state of both atoms is finally required when they exit the cavities.
     We also give a brief evaluation of the experimental feasibility of the scheme.
\end{abstract}

\section{Introduction}
Entanglement of separate systems certainly represents one of the
most striking features of quantum physics, both for its fundamental
nonlocal behavior \cite{ein,bell} and for its applications in
quantum information processing \cite{ben,pre}. For these reasons, it
is important to have implementable methods to generate entangled
states between separate systems in various contexts.

In cavity quantum electrodynamics (CQED) several schemes, exploiting
typical atom-cavity interactions, have been proposed to obtain, in
two separate single-mode cavities, entangled one-photon number
states \cite{bergou,mey,mes,com,brow}. In this context, it would be
of interest to get entanglement between electromagnetic field states
with mesoscopic characteristics and therefore having a maximum
number of photons larger than one, so that the classical-quantum
border may be investigated.

In this paper we propose a conditional scheme aimed at generating
entangled two-photons generalized binomial states (2GBSs) inside two
separate single-mode high-$Q$ cavities. Our scheme requires that the
two cavities are initially prepared in entangled one-photon
generalized binomial states \cite{lof}. In order to reach our goal,
we exploit the passage of two appropriately prepared atoms one in
each cavity and measure the atomic state of both atoms when they
exit the cavities. We will show that the probability of success is
larger than or equal to $1/2$, and depends on the degree of
entanglement we wish establish between the two cavities. We also
give a brief evaluation of the implementation of the scheme.

This paper is organized as follows: in Sec. \ref{gensch} we recall
the Jaynes-Cummings model that regulates the dynamics of the system
and we illustrate the scheme to generate entangled 2GBSs in two
separate cavities, briefly discussing its implementation; in Sec.
\ref{concl} we summarize our conclusions.

\section{Generation scheme\label{gensch}}
The generation scheme here proposed exploits the resonant
interaction of a two-level atoms with a single-mode high-$Q$ cavity
described by the usual Jaynes-Cummings Hamiltonian \cite{jay}
\begin{equation}
H_{JC}=\hbar\omega\sigma_{z}/2+\hbar\omega a^{\dag}a+i\hbar
g(\sigma_{+}a-\sigma_{-}a^{\dag}) \label{H}
\end{equation}
where $g$ is the atom-field coupling constant, $\omega$ is the
resonant cavity field mode, $a$ and $a^{\dag}$ are the field
annihilation and creation operators and
$\sigma_{z}=\ket{\uparrow}\bra{\uparrow}-\ket{\downarrow}\bra{\downarrow}$,
$\sigma_{+}=\ket{\uparrow}\bra{\downarrow}$,
$\sigma_{-}=(\sigma_{+})^{\dag}$ are the pseudo-spin operators,
$\ket{\uparrow}$ and $\ket{\downarrow}$ being respectively the
excited and ground state of the two-level atom. It is well known
that the Hamiltonian $H_{JC}$ generates the evolutions
\cite{mey,comp}
\begin{eqnarray}
\ket{\uparrow}\ket{n}\equiv\ket{\uparrow,n}&\rightarrow&\cos(g\sqrt{n+1}t)\ket{\uparrow,n}-\sin(g\sqrt{n+1}t)\ket{\downarrow,n+1}\nonumber\\
\ket{\downarrow}\ket{n}\equiv\ket{\downarrow,n}&\rightarrow&\cos(g\sqrt{n}t)\ket{\downarrow,n}+\sin(g\sqrt{n}t)\ket{\uparrow,n-1},\label{evo}
\end{eqnarray}
where $a^\dag a\ket{n}=n\ket{n}$.

\begin{figure}
\begin{center}
\includegraphics[width=0.6\textwidth, height=0.22\textheight]{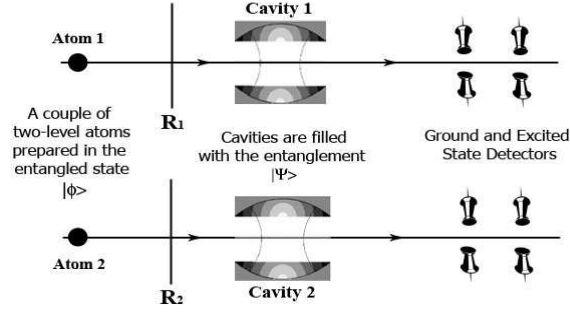}
\caption{\label{scheme}Experimental scheme for the generation of
entangled two-photons binomial states in two separate
cavities.}\end{center}
\end{figure}

\begin{figure}
\begin{center}
\includegraphics[width=0.6\textwidth, height=0.25\textheight]{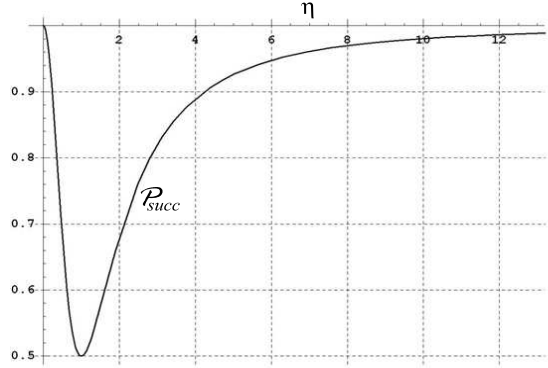}
\caption{\label{probsucc}Plot of the probability
$\mathcal{P}_{succ}$ to obtain entangled two-photons binomial states
in function of the parameter of entanglement
$|\eta|\in[0,+\infty[$.}\end{center}
\end{figure}

The scheme we are going to discuss, illustrated for simplicity in
Fig.\ref{scheme}, follows standard procedures currently used for
CQED experiments \cite{har,har1}.
We suppose that the two cavities are initially prepared in entangled
orthogonal one-photon generalized binomial states given by
\cite{lof}
\begin{equation}
\ket{\Psi^{(1)}}=\mathcal{N_{\eta}}\Big[\ket{p_1,\vartheta_1}_1\ket{1-p_2,\pi+\vartheta_2}_2
+\eta\ket{1-p_1,\pi+\vartheta_1}_1\ket{p_2,\vartheta_2}_2\Big],\label{entbernoulli}
\end{equation}
where $\eta\in\mathbb{R}$, ${\cal
N}_\eta=1/\sqrt{1+\modul{\eta}^{2}}$ is a normalization constant and
the state
\begin{equation}
\ket{p,\vartheta}=\sqrt{1-p}\ket{0}+e^{i\vartheta}\sqrt{p}\ket{1}\label{bernoulli}
\end{equation}
represents the one-photon generalized binomial state with a
probability of a single photon occurrence $p\in]0,1[$ and mean phase
$\vartheta$ \cite{sto,vid}. In Eq.~(\ref{entbernoulli}),
$\ket{p_j,\vartheta_j}_j$ denotes the single-mode state of the
$j$-th cavity. The state $\ket{\Psi^{(1)}}$ can be deterministically
obtained following, for example, the scheme reported in
Ref.~\cite{lof}. Consider now a couple of identical two-level
Rydberg atoms in the entangled state
\begin{equation}
\ket{\phi}={\cal
N}_\eta(\ket{\uparrow_1\downarrow_2}-\eta\ket{\downarrow_1\uparrow_2}).\label{entat12}
\end{equation}
This state can be prepared using, for example, the scheme suggested
by Gerry \cite{gerry}. Before entering the cavity~$j$, the $j$-th
atom crosses a Ramsey zone $R_j$ ($j=1,2$) where it resonantly
interacts with a classical field undergoing the transformations
\begin{eqnarray}
\ket{\uparrow_j}&\rightarrow&\ket{\uparrow}_{\mathbf{u}_j}\equiv\cos(\theta_j/2)\ket{\uparrow_j}-e^{i\varphi_j}\sin(\theta_j/2)\ket{\downarrow_j}\nonumber\\
\ket{\downarrow_j}&\rightarrow&\ket{\downarrow}_{\mathbf{u}_j}\equiv
e^{-i\varphi_j}\sin(\theta_j/2)\ket{\uparrow_j}+\cos(\theta_j/2)\ket{\downarrow_j},
\label{ramsequ}
\end{eqnarray}
with the versor
$\mathbf{u}\equiv(-\sin\theta\cos\varphi,-\sin\theta\sin\varphi,\cos\theta)$.
As well known, by adjusting the amplitude of the classical Ramsey
field as well as the interaction time between this field and the
atom, we have the possibility to control both the two quantities
$\theta$, the so-called ``Ramsey pulse'', and $\varphi$. Stated
another way, we can prepare an arbitrary superposition of the two
atomic states $\ket{\uparrow_j},\ket{\downarrow_j}$. Putting in
particular
\begin{equation}
\cos(\theta_j/2)\equiv\sqrt{p_j},\
\sin(\theta_j/2)\equiv\sqrt{1-p_j};\
\varphi_j=-\vartheta_j\label{cosp}
\end{equation}
where $p_1,p_2, \vartheta_1,\vartheta_2$ are the same parameters
that appear in the entangled state $\ket{\Psi^{(1)}}$ of
Eq.~(\ref{entbernoulli}), the transformations of
Eqs.~(\ref{ramsequ}) can be written as
\begin{eqnarray}
\ket{\uparrow_j}&\rightarrow
&\ket{\uparrow}_{\mathbf{u}_j}\equiv\sqrt{p_j}\ket{\uparrow_j}-e^{-i\vartheta_j}\sqrt{1-p_j}\ket{\downarrow_j}\nonumber\\
\ket{\downarrow_j}&\rightarrow
&\ket{\downarrow}_{\mathbf{u}_j}\equiv
e^{i\vartheta_j}\sqrt{1-p_j}\ket{\uparrow_j}+\sqrt{p_j}\ket{\downarrow_j}.\label{ramseqj}
\end{eqnarray}
Therefore, the total atom-cavity state after the Ramsey zones is
\begin{equation}
\ket{\Phi(0)}=\mathcal{N}_{\eta}\big(\ket{\uparrow}_{\mathbf{u}_1}\ket{\downarrow}_{\mathbf{u}_2}-
\eta\ket{\downarrow}_{\mathbf{u}_1}\ket{\uparrow}_{\mathbf{u}_2}\big)\ket{\Psi^{(1)}}.\label{Phi0}
\end{equation}
Successively each atom enters the respective cavity where it
resonantly interacts with the single-mode electromagnetic field. The
dynamics of the two subsystems ``atom+cavity'' 1 and 2 are
independent. So, starting from the total atom-cavity state of
Eq.~(\ref{Phi0}) and using the explicit expressions of the cavity
and atomic states given in Eqs.~(\ref{entbernoulli}),
(\ref{bernoulli}), (\ref{ramseqj}), we exploit the Jaynes-Cummings
evolutions as given by Eq.~(\ref{evo}) in correspondence to each
subsystem $j=1,2$. Indicate by $T_j$ the interaction time between
the atom $j$ and the $j$-th cavity, and suppose that the conditions
\begin{equation}
\sin gT_j+\cos
gT_j=\sqrt{2},\quad\sin(g\sqrt{2}T_j)=1\label{sincond}
\end{equation}
are satisfied. Under these hypothesis we obtain
\begin{eqnarray}
\ket{\uparrow}_{\mathbf{u}_j}\ket{p_j,\vartheta_j}_j&\stackrel{T_j}{\rightarrow}
&-e^{-i\vartheta_j}\ket{2,p_j,\vartheta_j}_j\ket{\downarrow_j},\nonumber\\
\ket{\downarrow}_{\mathbf{u}_j}\ket{p_j,\vartheta_j}_j&\stackrel{T_j}{\rightarrow}&
\frac{1}{\sqrt{2}}\big[e^{i\vartheta_j}\ket{0_j\uparrow_j}+\ket{\Gamma_j(p_j,\vartheta_j)}\big],\nonumber\\
\ket{\uparrow}_{\mathbf{u}_j}\ket{1-p_j,\pi+\vartheta_j}_j&\stackrel{T_j}{\rightarrow}&
\frac{1}{\sqrt{2}}\big[\ket{0_j\uparrow_j}-e^{-i\vartheta_j}\ket{\Gamma_j(p_j,\vartheta_j)}\big],\nonumber\\
\ket{\downarrow}_{\mathbf{u}_j}\ket{1-p_j,\pi+\vartheta_j}_j&\stackrel{T_j}{\rightarrow}
&\ket{2,1-p_j,\pi+\vartheta_j}_j\ket{\downarrow_j},\label{binevo}
\end{eqnarray}
where we have indicated with $\ket{2,p,\vartheta}$ the generalized
binomial state with a maximum number of photons $N=2$, probability
of a single photon occurrence $p$ and mean phase $\vartheta$,
defined as \cite{sto,vid}
\begin{equation}
\ket{2,p,\vartheta}\equiv\sum_{n=0}^2\left[{2\choose
n}p^{n}(1-p)^{2-n}\right]^{1/2}e^{in\vartheta}\ket{n},\label{bin2}
\end{equation}
and where we have set
\begin{equation}
\ket{\Gamma_j(p_j,\vartheta_j)}\equiv\sqrt{2p_j(1-p_j)}\ket{0_j}-(1-2p_j)e^{i\vartheta_j}\ket{1_j}
-\sqrt{2p_j(1-p_j)}e^{2i\vartheta_j}\ket{2_j}.
\end{equation}
Unfortunately the conditions of Eq.~(\ref{sincond}) cannot be
simultaneously satisfied. Let us however observe that, solving the
first condition of Eq.~(\ref{sincond}), we get $gT_j=\pi/4+2m_j\pi$,
where $m_j$ is a non-negative integer. Thus we can look for suitable
values of $m_j$ in correspondence of which the second condition of
Eq.~(\ref{sincond}) is approximatively satisfied. The choice of
$m_j$ however must be done coherently with the typical experimental
values of the interaction times in CQED systems
($gT_j\sim10^{-1}\div10^2$ \cite{har}), thus confining the values of
$m_j$ inside the range $0\leq m_j\leq16$. Fixing, in particular,
$m_j=5$ and therefore the same interaction time in the two cavities
given by
\begin{equation}
T_1=T_2=T=41\pi/4g,\label{T}
\end{equation}
we find that
\begin{equation}
\sin(g\sqrt{2}T)=1-\delta,\ \textrm{where
$\delta\sim10^{-4}$}.\label{sinap}
\end{equation}
Within this approximation we can consider the second condition of
Eq.~(\ref{sincond}) as satisfied, too. So, utilizing
Eq.~(\ref{binevo}) with $T_j=T$, we find that, if the condition
$\vartheta_1=\vartheta_2\equiv\vartheta$ is verified, the total
atom-cavity state $\ket{\Phi(0)}$ of Eq.~(\ref{Phi0}) evolves, apart
from a global phase factor, into
{\setlength\arraycolsep{2pt}\begin{eqnarray}
\ket{\Phi(T)}&=&\mathcal{N}_\eta^2\big\{\big[\ket{2,p_1,\vartheta}_1\ket{2,1-p_2,\pi+\vartheta}_2
-\eta^2\ket{2,1-p_1,\pi+\vartheta}_1\ket{2,p_2,\vartheta}_2\big]\ket{\downarrow_1\downarrow_2}\nonumber\\
&-&\eta
e^{i\vartheta}\ket{0_1}\ket{\Gamma_2(p_2,\vartheta)}\ket{\uparrow_1\downarrow_2}
+\eta
e^{i\vartheta}\ket{\Gamma_1(p_1,\vartheta)}\ket{0_2}\ket{\downarrow_1\uparrow_2}\big\}.\label{enttot}
\end{eqnarray}}From Eq.~(\ref{enttot}), it is readily noticed that, if both atoms
are measured in the ground state $\ket{\downarrow}$ after exiting
the cavities, the resulting two-cavities field state is
\begin{equation}
\ket{\Psi^{(2)}}=\mathcal{N}_2\big[\ket{2,p_1,\vartheta}_1\ket{2,1-p_2,\pi+\vartheta}_2
-\eta^2\ket{2,1-p_1,\pi+\vartheta}_1\ket{2,p_2,\vartheta}_2\big].\label{entbin2}
\end{equation}
Since it is satisfied the orthogonality condition
$\scal{2,p,\vartheta}{2,1-p,\pi+\vartheta}=0$ \cite{lof}, the
normalization constant has the value
$\mathcal{N}_2=1/\sqrt{1+\modul{\eta}^4}$, and the state
$\ket{\Psi^{(2)}}$ of Eq.~(\ref{entbin2}) represents entangled
orthogonal 2GBSs in two separate single-mode high-$Q$ cavities. From
Eq.~(\ref{enttot}), the probability of finding both atoms in the
ground state after exiting the cavities, i.e. the probability of
success to obtain our target state $\ket{\Psi^{(2)}}$, is given by
\begin{equation}
\mathcal{P}_{succ}=\modul{\bra{\Psi^{(2)}}\bra{\downarrow_1\downarrow_2}\Phi(T)\rangle}^2=
\mathcal{N}_{\eta}^4/\mathcal{N}_2^2\Rightarrow\mathcal{P}_{succ}=(1+\modul{\eta}^4)/(1+\modul{\eta}^2)^2.\label{Psucc}
\end{equation}
Following Ref.~\cite{abo}, the degree of entanglement of the state
$\ket{\Psi^{(2)}}$ is given by
$G^{(E)}=2\modul{\eta}^2/(1+\modul{\eta}^4)$ and it is invariant
with respect to the substitution
$\modul{\eta}\rightarrow1/\modul{\eta}$, equal to zero for
$\modul{\eta}=0,+\infty$ (uncorrelated states) and equal to one for
$\modul{\eta}=1$ (maximally entangled states). In terms of
$G^{(E)}$, the probability of success $\mathcal{P}_{succ}$ given in
Eq.~(\ref{Psucc}) becomes
$\mathcal{P}_{succ}=1/\big(1+G^{(E)}\big)$. In Fig.\ref{probsucc} we
plot the graph of $\mathcal{P}_{succ}$ versus the parameter of
entanglement $|\eta|$.
The probability of success to obtain maximally entangled 2GBSs, when
$|\eta|=G^{(E)}=1$, has its minimum value
$\mathcal{P}_{succ}^{(min)}=1/2$ and it tends to one when
$|\eta|\rightarrow0,+\infty$, i.e. when the degree of entanglement
$G^{(E)}$ tends to zero. The probability of success is equal to one
for $|\eta|=0,+\infty$, but in this case, with $G^{(E)}=0$, there is
no entanglement at all.

As remarked, the state $\ket{\Psi^{(2)}}$ of Eq.~(\ref{entbin2})
represents entangled two-photons generalized binomial states.
However, when the probabilities of a single photon occurrence $p_j$
($j=1,2$) take their limit values $p_j=0,1$, the state
$\ket{\Psi^{(2)}}$ reduces to entangled number states with zero or
two photons inside the cavities. In fact, using the property that a
binomial state with a maximum number of photons $N$ is equal to
$\ket{0}$ for $p=0$ and to $\ket{N}$ for $p=1$ \cite{sto}, from
Eq.~(\ref{entbin2}) we obtain
\begin{eqnarray}
\ket{\Psi^{(2)}}_{p_1=p_2=1}&=&\mathcal{N}_2\big[\ket{2_10_2}-\eta^2\ket{0_12_2}\big],\nonumber\\
\ket{\Psi^{(2)}}_{p_1=1,p_2=0}&=&\mathcal{N}_2\big[\ket{2_12_2}-\eta^{2}
e^{-4i\vartheta}\ket{0_10_2}\big].
\end{eqnarray}

We shall give here a brief evaluation of some experimental errors
involved in the implementation of our generation scheme. A necessary
condition required by this scheme is that the atoms cross the
cavities for a predeterminate time $T$. The experimental
uncertainties on selected velocity and interaction time, $\Delta v$
and $\Delta T$, are such that $\Delta T/T\approx\Delta v/v$. In
current laboratory experiments we have $\Delta v/v\sim10^{-2}$ or
less \cite{har1,hag}. It is also possible to see that the error
$\delta$ of Eq.~(\ref{sinap}) is much smaller than the error induced
by these experimental values of $\Delta T/T$ on the condition
$\sin(g\sqrt{2}T)=1$. Another aspect we have ignored is the atomic
or photon decay during the atom-cavity interactions. This assumption
is valid if $\tau_{at},\tau_{cav}>T$, where $\tau_{at},\tau_{cav}$
are the atomic and photon mean lifetimes respectively and $T$ is the
interaction time. For Rydberg atomic levels and microwave
superconducting cavities with quality factor $Q\sim10^8\div10^{10}$,
we have $\tau_{at}\sim10^{-5}\div10^{-2}\textrm{s}$ and
$\tau_{cav}\sim10^{-4}\div10^{-1}\textrm{s}$. Since typical
atom-cavity field interaction times are
$T\sim10^{-5}\div10^{-4}\textrm{s}$, the required condition on the
mean lifetimes can be satisfied \cite{har}. Moreover, the typical
mean lifetimes of the Rydberg atomic levels $\tau_{at}$ must be such
that the atoms do not decay during the entire sequence of the scheme
and, since the proposed scheme requires that the cavities are
initially prepared in entangled one-photon generalized binomial
states \cite{lof}, the photon mean lifetimes $\tau_{cav}$ must be
long enough to permit cavity fields not to decay before they
interact with the successive atoms \cite{har,har1}. Finally, we
should consider the fact that experimental detectors efficiencies
are smaller than one, so we could have no ``click'' when an atom
crosses the field ionization detector: in this case the generation
scheme should be repeated from the beginning.

\section{Conclusion\label{concl}}
In this paper we have proposed a conditional scheme for the
generation of entangled two-photons generalized binomial states in
two spatially separate single-mode high-$Q$ cavities. This scheme
exploits standard atom-cavity interactions and requires the final
measurement of the atomic states. The probability of success to
generate our target state is always larger than or equal to $1/2$,
depending on the value of the parameter of entanglement $|\eta|$ and
therefore on the degree of entanglement $G^{(E)}$. In particular, we
have seen that the probability of success to obtain maximally
entangled two-photons binomial states, i.e. when $|\eta|=G^{(E)}=1$,
takes its minimum value $\mathcal{P}_{succ}^{(min)}=1/2$ and it
tends to one when $G^{(E)}$ tends to zero.

Finally, we have briefly estimated the typical experimental errors
involved in such CQED systems, and we have seen that our generation
scheme is not sensibly affected by these errors. This shows that the
implementation of our generation scheme is within the current
experimental technics \cite{har1}.

As far as we know, the scheme proposed here for the generation of
entangled two-photons generalized binomial states in two separate
cavities represents the first example, in the context of the CQED,
of a scheme that permits to produce entanglement between
non-classical states of the electromagnetic field having non zero
mean fields and a number of photons greater than one. In conclusion,
this kind of entangled state could be useful for fundamental
investigations on nonlocal properties, for studying field
correlations between the cavities or Bell's inequality violations,
and for applications in quantum computation and information
processing.

\end{document}